\documentclass[twocolumn,pra,showpacs,superscriptaddress,amssymb,amsmath,amsmath,floatfix]{revtex4-2}
\usepackage[T1]{fontenc}
\usepackage{color,graphicx}
\usepackage{epstopdf}
\usepackage{bm}
\usepackage[colorlinks=true,linkcolor=blue,urlcolor=blue,citecolor=blue]{hyperref}
\graphicspath{{./figs/}}
\usepackage{float}
\usepackage{times}
\usepackage{amsmath,amsthm,amssymb,amsfonts}
\usepackage{xcolor}
\usepackage{orcidlink}

\newcommand{\be}{\begin{equation}}
\newcommand{\ee}{\end{equation}}

\begin{document}
\title{Effects of interaction range on the mean-field dynamics of Bose polarons}

\author{Piotr Wysocki\orcidlink{0009-0009-1589-1524}
}
\affiliation{Institute for Quantum Optics and Quantum Information of the Austrian Academy of Sciences, 6020 Innsbruck, Austria}
\affiliation{Institute for Theoretical Physics, University of Innsbruck, 6020 Innsbruck, Austria}
\affiliation{Faculty of Physics, University of Warsaw, Pasteura 5, 02-093 Warsaw, Poland}
\author{Ubaldo Cavazos Olivas\orcidlink{0000-0003-4052-4717}
}
\affiliation{Faculty of Physics, University of Warsaw, Pasteura 5, 02-093 Warsaw, Poland}
\author{Marek Tylutki\orcidlink{0000-0002-4243-3803}
}
\email{marek.tylutki@pwr.edu.pl}
\affiliation{Institute of Theoretical Physics, Wroc{\l}aw University of Science and Technology, 50-370 Wroc{\l}aw, Poland}
\author{Krzysztof Jachymski\orcidlink{0000-0002-9080-0989}
}
\email{krzysztof.jachymski@fuw.edu.pl}
\affiliation{Faculty of Physics, University of Warsaw, Pasteura 5, 02-093 Warsaw, Poland}

\date{\today}

\begin{abstract}
We consider the three-dimensional Bose polaron problem in the regime of finite range interactions and competing length scales. Working in the reference frame of the impurity, we study both static and out of equilibrium properties of the system, in particular the transfer of momentum between the impurity and the host gas. We find that relaxation dynamics can occur via damped oscillations of the impurity velocity with simple dependence on the interaction strength. Furthermore, the equilibration process is sensitive to the type of the impurity-bath interaction. Specifically, interatomic forces describing ion-atom systems lead to much longer timescales and more pronounced oscillations in the strong coupling regime with respect to local interaction potentials. We also find that the effective masses can differ by a large amount between the two scenarios, even if the number of atoms in the  polaron cloud remains similar for both cases.

\end{abstract}

\maketitle
\section{Introduction}
Dynamics of an impurity moving in an interacting medium is a paradigmatic problem in modern physics, and it is highly relevant for our understanding of the properties of matter. While being a complex many-body problem, it encourages approximate treatments and can be seen as a testbed for simplified models. Examples cover a wide range of scales from Brownian motion of particles in classical fluids to strongly interacting quantum degenerate systems such as neutron stars. In quantum mechanics, the notion of quasiparticles offers a radical simplification of the problem, wherein the impurity becomes dressed by elementary excitations of the medium and forms a polaron~\cite{Landau1948}. Even simplistic variational wave function involving a single Bogoliubov phonon can give surprisingly accurate predictions for the ground state energy. 

\begin{figure}
 \includegraphics[width = .98\columnwidth]{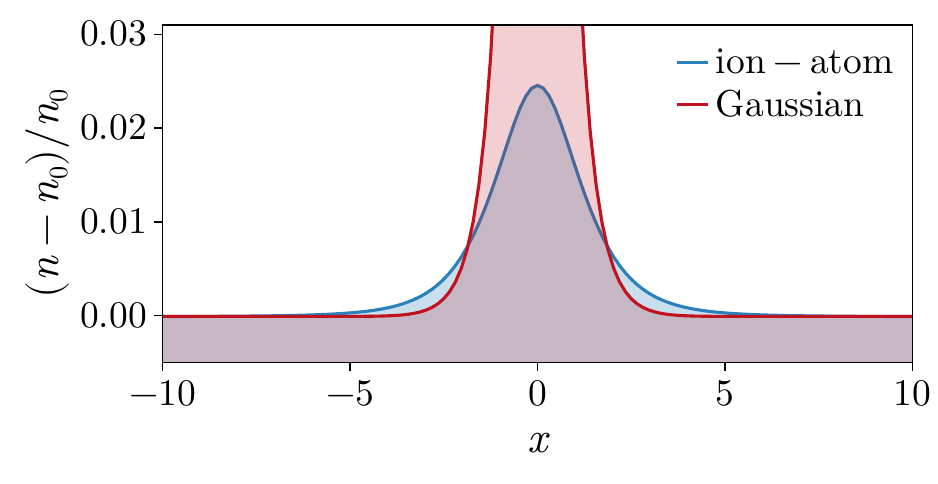}
	\caption{\label{fig:ground} Ground state density profile for a BEC interacting with a charged impurity vs a gaussian one for similar impurity-boson scattering length.}
\end{figure}
Ultracold atomic systems provide a playground for exploring various aspects of polaron physics. This is due to their excellent microscopic understanding as well as a high degree of controllability of virtually all system parameters. In recent years, both Fermi and Bose polarons were realized in experiments~\cite{Schirotzek2009,Cetina2016,Ness2020,Hu2016,Yan2020,Yan2024,Etrych2025}, typically by mixing two atomic species, and they were studied across a wide range of the impurity-bath coupling strengths by means of Feshbach resonances. They were also studied in semiconducting materials~\cite{Tan2023} (for a comprehensive review, see~\cite{Grusdt2024,Massignan2025}). Powerful theoretical approaches~\cite{Christensen2015,Shchadilova2016,Christianen2024} as well as numerical studies~\cite{Ardila2015,Mistakidis2019} have been developed to characterize these systems and predict not only the energy, but also the whole spectral function, quasiparticle residue as well as the effective mass. So far, less attention has been paid to the dynamics of polarons that are moving through the medium with finite speed~\cite{Skou2021}. In the case of superfluids, which are naturally realized in Bose gases, the Landau theory introduces the notion of critical velocity which, for linear dispersion, equals the speed of sound and under which the flow is dissipationless. Under these circumstances, the drag force acting on the impurity should vanish. However, this result comes from perturbative treatment based on the analysis of Bogoliubov excitation spectrum, and it is well known that for strong impurity-bath interactions the drag appears already below the speed of sound~\cite{Pavloff2002,Ianeselli2006,Carusotto2006}. Similarly, using a variational treatment based on a gaussian state of Bogoliubov excitations it was predicted that initializing the impurity at high momenta would lead to emission of an analog of Cherenkov radiation and fast relaxation to the steady state~\cite{Seetharam2021,Seetharam2024}. However, accounting for the change of the gas density around the impurity can be expected to turn this sharp transition into a crossover, as now the speed of sound would depend on the distance from the impurity.

In typical experimental settings the ultracold gas is extremely dilute and mean interparticle distance is much larger than the interaction range. This validates the use of pseudopotentials, simplifying the model significantly. Also long-range interactions can be realized; for instance using charged impurities~\cite{Tomza2019,Lous2022}. This leads to modifications of the properties of the system~\cite{Massignan2005,Christensen2021,Astrakharchik2021,Cavazos2024,Yogurt2025,Sanchez2026}. Simple approach allowing to study various interaction types relies on the mean-field theory developed in the impurity's reference frame, i.e. after performing the Lee-Low-Pines transformation~\cite{Gross1962,Drescher2020,Schmidt2022,Yegovtsev2024}. This type of treatment allows to get insight into the dynamics, such as an experimentally realistic quench scenario involving creation of the impurity at finite momentum in an initially homogeneous gas. Interestingly, for strong enough interactions, the model predicts oscillations of the impurity momentum, which can even bounce back with respect to the original direction of its initial momentum for some time. This effect occurs both in one and three dimensions~\cite{Majumdar2025,Majumdar2026,Wysocki2026}, which makes it distinct from the quantum flutter phenomenon present in strongly correlated one-dimensonal media~\cite{Mathy2012,Knap2014,Burovski2014,Zhang2024}.

It is well known that dilute gases in the ultracold regime can be described using zero-range pseudopotentials with the coupling strength proportional to the scattering length. Finite energy corrections in the form of effective range can be incorporated as well~\cite{Collin2007,Schmidt2022}. In this work, we study how the range of the interactions can manifest in the properties of Bose polarons beyond the universal regime, see Fig.~\ref{fig:ground} for the corresponding density profiles, comparing a short-range interaction modelled by a gaussian potential with a potential of a charged particle that exhibits power-law decay, naturally occurring for ionic impurities. As the scattering length is no longer the key quantity describing the potential, the parameter space is large. We show that while some properties of the system such as the effective mass of the polaron can be tuned to be similar in both cases, strong differences always prevail in the impurity dynamics. 

This work is structured as follows. We introduce the system and briefly describe our treatment in Section~\ref{sec:system}. Then in Sec.~\ref{sec:res} we discuss both static and dynamic properties of long- and short-range polarons. Section~\ref{sec:concl} concludes the paper.

\section{The system}
\label{sec:system}

\subsection{Hamiltonian and the mean-field approximation}
The system consists of a single impurity interacting with a degenerate three-dimensional Bose gas at zero temperature. The Hamiltonian reads
\begin{equation}
	\label{eq.mbham}
	\mathcal{H} = \frac{{\bm p}_I^2}{2m_I}+\sum_i \frac{{\bm p}_i^2}{2m_B}+\sum_i V({\bm x}_i - {\bm x}_I) + \sum_{i<j} U({\bm x}_i - {\bm x}_j) ~,
\end{equation}
where ${\bm p}_i$ are the momenta of the bosons, ${\bm p}_I$ is the momentum of the impurity, and $m_B$, $m_I$ are the atomic masses of the respective species, $U({\bm x}_i - {\bm x}_j)$ denotes the interaction potential within the bath and $V({\bm x}_i - {\bm x}_I)$ the interaction potential between the impurity and the gas. We consider two types of interaction potentials: (i) a power-law potential with short-range regularization~\cite{Krych2015}
\begin{equation}
	\label{eq.ionpot}
	V({\bm x}) = -\frac{C_4}{({\bm x}^2 + b^2)^2} ~,
\end{equation}
which is suitable for charged impurities, and (ii) a rapidly decaying, finite-ranged gaussian potential
\begin{equation}
	\label{eq.gausspot}
	V({\bm x}) = -V_0 e^{-{\bm x}^2/\sigma^2}\, .
\end{equation}
The bosonic interactions are described by the standard pseudopotential  $g\delta(\mathbf{r}_i-\mathbf{r}_j)$.
Following previous works, we apply the Lee-Low-Pines (LLP) transformation~\cite{LLP,Girardeau1961,Drescher2020} using $S = {\bm x}_I \cdot \sum_i {\bm p}_i$ to take advantage of the conservation of the total momentum. Note that the LLP transformation assumes a Galilean invariance of the system, which must be preserved in subsequent numerical calculations in order to avoid unphysical results~\cite{Jager2020,Breu2025}. Now, we introduce a macroscopic wave function $\psi({\bm x}, t)$ and perform a mean-field decoupling for bosons, demanding the gas parameter to remain small everywhere. The resulting enegy functional reads
\begin{equation} \label{eq.edf}
	\mathcal{E}[\psi] = \mathcal{E}_0[\psi] + \frac{({\bm p}_0 - \langle {\bm P} \rangle)^2}{2m_I}\\
\end{equation}
where $\mathcal{E}_0$ is the energy of the standard Gross-Pitaevskii functional 
\begin{equation*}
	\mathcal{E}_0[\psi] = \int d{\bm x} \left( \frac{\hbar^2}{2m_r}| \nabla \psi|^2 + V({\bm x}) |\psi|^2 + \frac{g}{2} |\psi|^4 \right)
\end{equation*}
with the reduced mass $m_r = m_I m_b / (m_I + m_b)$ and $\langle {\bm P} \rangle$ is the condensate's momentum. 
The variation of Eq.~(\ref{eq.edf}) with respect to the condensate's wave function leads to the modified Gross-Pitaevskii equation~\cite{Drescher2020,Majumdar2025,Wysocki2026}
\begin{equation}\label{eq:gpe}
	i \hbar \frac{\partial \psi}{\partial t} = -\frac{\hbar^2}{2m_r} \nabla^2 \psi + V({\bm x}) \psi + g |\psi|^2 \psi + \frac{i\hbar}{m_I} {\bm p}_I \cdot \nabla \psi 
\end{equation}
where the momentum of the impurity  ${\bm p}_I$ itself depends on the condensate wave function
\begin{equation*}
	{\bm p}_I(t) = {\bm p}_0 + i \hbar \int d{\bm x} \, \psi({\bm x}, t)^* \nabla \psi({\bm x}, t) ~.
\end{equation*}
It is important to mention that even though we use a mean-field ansatz, this approach in fact includes some impurity-medium correlation due to the form of the Lee-Low-Pines transformation, meaning that in the laboratory frame the wave function is entangled.

\begin{figure}
	\includegraphics[width = .98\columnwidth]{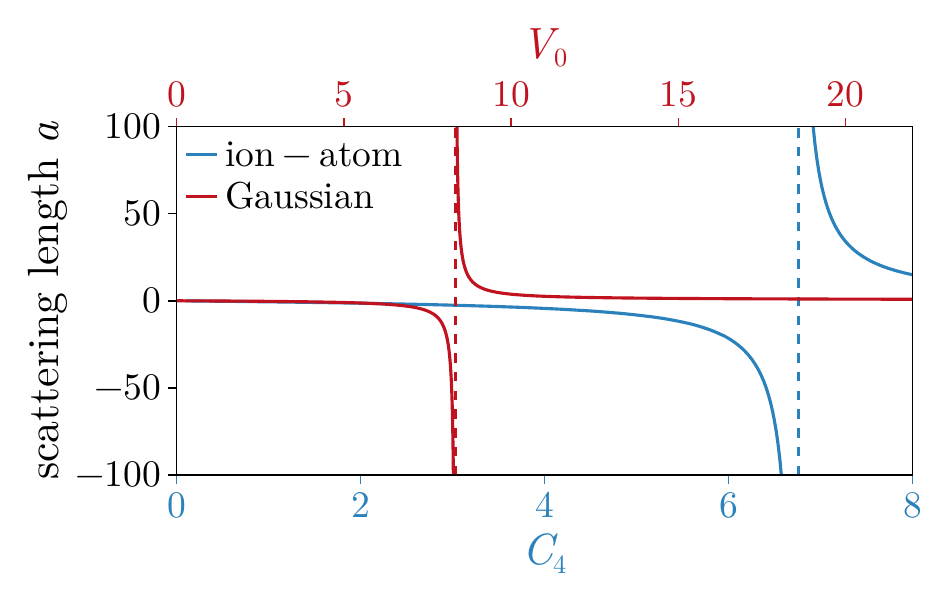}
	\caption{\label{fig:scatl} Scattering length $a_{IB}$ as a function of the interaction strength for the ion-atom and gaussian potentials with $b=1.5$ and $\sigma=0.4$, respectively (cf.~eq.~\eqref{eq.ionpot}-\eqref{eq.gausspot}). }
\end{figure}
Since we intend to compare two different types of interactions, it is convenient to set the units using the condensate healing length $\xi = \hbar/\sqrt{m_b n_0 g}$ as the unit of length, with $n_0$ being the gas density at the system boundary and $\varepsilon = n_0g$ as the unit of energy. For the impurity dynamics, the associated unit of time will be $\tau=\hbar/n_0 g$, and the speed of sound is therefore $c = 1$. The length scales introduced by the interactions are $R^\star=(2m_r C_4/\hbar^2)^{1/2}$, $b$, and $\sigma$. For dilute gases and short range interactions, the relevant parameter is the scattering length, which we show for the two potentials in Fig.~\ref{fig:scatl}, varying the overall interaction strength $C_4$ or $V_0$ for the ion's potential and the gaussian potential, respectively. Resonances occur where new bound states appear at the threshold.

\subsection{Hydrodynamic equations and model of momentum dynamics}
We can represent our mean-field equations in the hydrodynamic formulation, with $\psi({\bm x}, t) = \sqrt{n({\bm x}, t)} e^{i S({\bm x}, t)}$ and the superfluid velocity ${\bm v}({\bm x}, t) = \hbar \nabla S({\bm x}, t) / m$. With the impurity velocity defined as ${\bm v}_I = {\bm p}_I / m_I$, the boosted continuity equation takes the form
\begin{equation}
\frac{\partial n}{\partial t} = -\nabla [n ({\bm v} - {\bm v}_I)] ~,
\label{eq.hydro1}
\end{equation}
while the Hamilton-Jacobi equation for the phase $S$ in the reference frameof the impurity  reads
\begin{equation}
-\hbar \frac{\partial S}{\partial t} = \frac{m_r ({\bm v} - {\bm v}_I)^2}{2} - \frac{m_r {\bm v}_I^2}{2} + V({\bm x}) + gn + Q ~,
\label{eq.hydro2}
\end{equation}
with $Q = -\hbar^2 \nabla^2 \sqrt{n} / (2 m \sqrt{n})$ the quantum pressure term. As was shown, for instance in Ref.~\cite{Wysocki2026}, the equation for the impurity momentum is
\begin{equation}
\frac{d {\bm p}_I}{dt} =  \int d{\bm x} \ n \nabla V =  \int d{\bm x} \ \delta n \nabla V ~,
\end{equation}
which closes this system of equations. We can already see that an initial perturbation of ${\bm p}_I$ causes a delayed response of ${\bm p}_I$ at later times mediated by the excitation of $n({\bm x})$, a fact that we will make use of later in this work. We can linearize these equations assuming $n({\bm x}, t) = n_0 + \delta n({\bm x}, t)$ and $S({\bm x}, t) = S_0 + \delta S({\bm x}, t)$, which leads to
\begin{equation}
\frac{\partial}{\partial t} \begin{pmatrix} \delta n\\ \delta S\\ \end{pmatrix} = \begin{pmatrix}
{\bm v}_I \cdot \nabla & -\frac{\hbar n_0 \nabla^2}{m_r}\\
\frac{\hbar^2 \nabla^2}{4 m_r n_0} - \frac{g}{\hbar} & {\bm v}_I \cdot \nabla\\
\end{pmatrix} \begin{pmatrix} \delta n\\ \delta S\\ \end{pmatrix}
\end{equation}
These equations reveal that the linear response of the gas density depends on the momentum of the impurity. The movement of the impurity affects the dynamics of the density perturbations $\delta n$, and the density in turn enters the equation for $d{\bm p}_I  / dt$ in a nonlocal way. Therefore, the initial perturbation of the value of the impurity momentum 
determines its dynamics with a certain time delay once the condensate degrees of freedom have been integrated out. In general, we can express this as a Langevin equation with a memory kernel $\mathcal{K}(t)$~\cite{Kubo2012}
\begin{equation}
\delta \dot{\bm p}_I(t) = - \int_0^t d t^\prime \mathcal{K}(t - t^\prime) \delta {\bm p}_I(t^\prime) \approx -\int_0^\infty d \tau \, \mathcal{K}(\tau) \delta {\bm p}_I(t - \tau)
\end{equation}
If the timescales for the impurity are longer that the scale related to $\mathcal{K}(\tau)$, we can extend the upper integration limit to infinity. Now, we can expand the right-hand side in powers of $\tau$ as
\begin{equation}
\delta \dot{\bm p}_I(t) \approx -\int_0^\infty d \tau \, \mathcal{K}(\tau) \left[ \delta {\bm p}_I(t) - \tau \delta \dot{\bm p}_I(t) + \frac{\tau^2}{2} \delta \ddot{\bm p}_I(t) + \ldots \right]
\end{equation}
The time derivatives of ${\bm p}_I$ do not depend on $\tau$. While microscopic calculation of $\mathcal{K}$ seems to be feasible, we only note that it should decay in $\tau$ such that each term can be separately integrated. Upon renaming the coefficients, the equation takes the form of a damped harmonic oscillator, characteristic for dissipative environments~\cite{Caldeira1981}
\begin{equation}
\delta \ddot{\bm p}_I(t) + 2 \lambda \delta \dot{\bm p}_I(t) + \omega_0^2 \delta {\bm p}_I(t) = 0 ~.
\end{equation}
The coefficients are positive, reflecting the stability of the system and the dissipative dynamics. The solution in the underdamped regime is
\begin{equation}
{\bm p}_I(t) = {\bm p}_I^{(\infty)} + {\bm p}_0 e^{-\lambda t}\cos \left( \sqrt{\omega_0^2 - \lambda^2} \, (t - \delta t) \right) ~,
\label{eq.dampedho}
\end{equation}
where we introduced an initial time delay $\delta t$ necessary for a build-up of the perturbation and onset of the oscillatory dynamics. Numerical verification of this prediction is one of the main goals of this work. For lower interaction strengths, the damping coefficient increases (see subsequent numerical results) and the impurity experiences overdamped dynamics and exponential decay of the velocity, suggesting that it probes a different spectral region of the bath of Bogoliubov excitations in this regime.

\section{Results}
\label{sec:res}
\subsection{Ground state}
\begin{figure}
\includegraphics[width = .98\columnwidth]{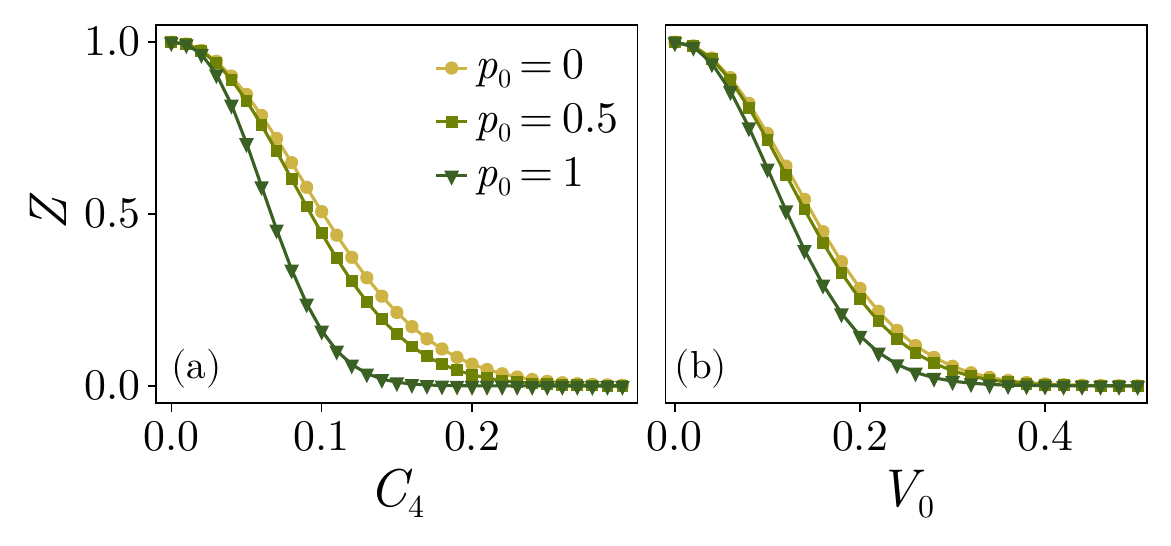}
\caption{\label{fig:residue} Polaron residue as a function of the interaction strength $C_4$ or $V_0$ for zero- as well as finite momentum ground states.}
\end{figure}
\begin{figure}
\includegraphics[width=0.4\textwidth]{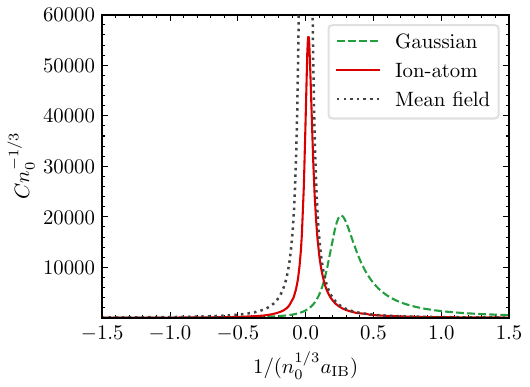} 
\caption{\label{fig:tan-contact}Tan contact as a function of the inverse scattering length for a weakly interacting impurity (gray dotted line), ion-atom (solid red line) and Gaussian (green-dashed curve) potentials.}
\end{figure}
\begin{figure}
\includegraphics[width = .98\columnwidth]{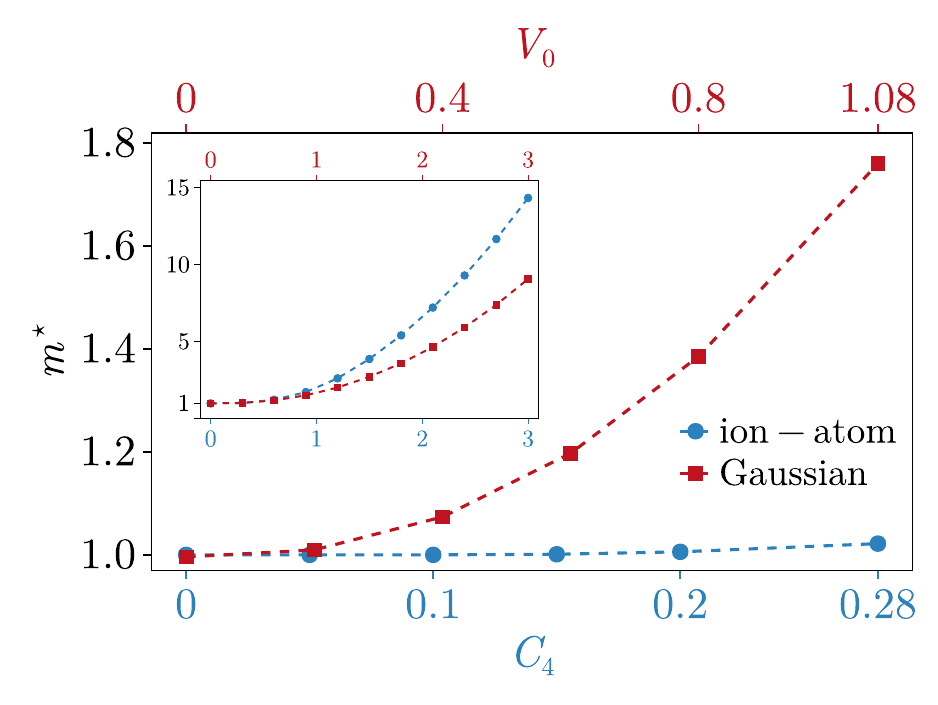}
\caption{\label{fig:meff} Effective masses for the ion-atom (upper blue curve) vs gaussian (lower red curve) potentials as a function of the interaction strength $C_4$ and $V_0$, respectively. The mass is given in units of the bare atomic mass. The scaling of potential strength is chosen such that similar number of atoms forms the polaronic cloud in both cases. Inset: same, but without scaling the $x$ axis.}
\end{figure}
First we look at the ground state solution, performing numerical calculations of the Gross-Pitaevskii equation in imaginary time at finite momentum ${\bm p}_0$. We vary the interactions parameters to get qualitatively similar static properties for both the ion-atom and gaussian potentials. This turns out to be rather subtle: at very weak coupling, the properties are expected to be universal and depend on the scattering length, but for finite deformation of the gas and increase in the gas parameter this is no longer the case. We found that, in particular, for the two potentials leading to similar peak gas density, the two resulting polarons strongly differ in the number of participating atoms. In the examples below, we settle for an intermediate scenario, for which the deformation of the gas density, quantified by the dressing cloud size defined as $\Delta N=\int{d\mathbf{r}(n(\mathbf{r})-n_0)}$, affects similar number of atoms, and the scattering length is also in the same regime. Specifically, choosing $C_4=0.28$ and $V_0=1.08$, we get $\Delta N_{\rm ia} / \Delta N_{\rm g} \approx 1.1$ and $a_{\rm ia} / a_{\rm g} \approx 1.5$.

The shape of the bosonic cloud obtained via imaginary time propagation for the parameters chosen above and at ${\bm p}_0 = 0$ is depicted in Figure~\ref{fig:ground}. We observe that the increase in peak density for the ion-atom potential is an order of magnitude lower than for the gaussian potential, which means that the even though the number of atoms in the cloud in both cases is similar, the density profiles are vastly different. This is expected due to the slower decay of the ion-atom potential. 

\begin{figure}
\includegraphics[width = .98\columnwidth]{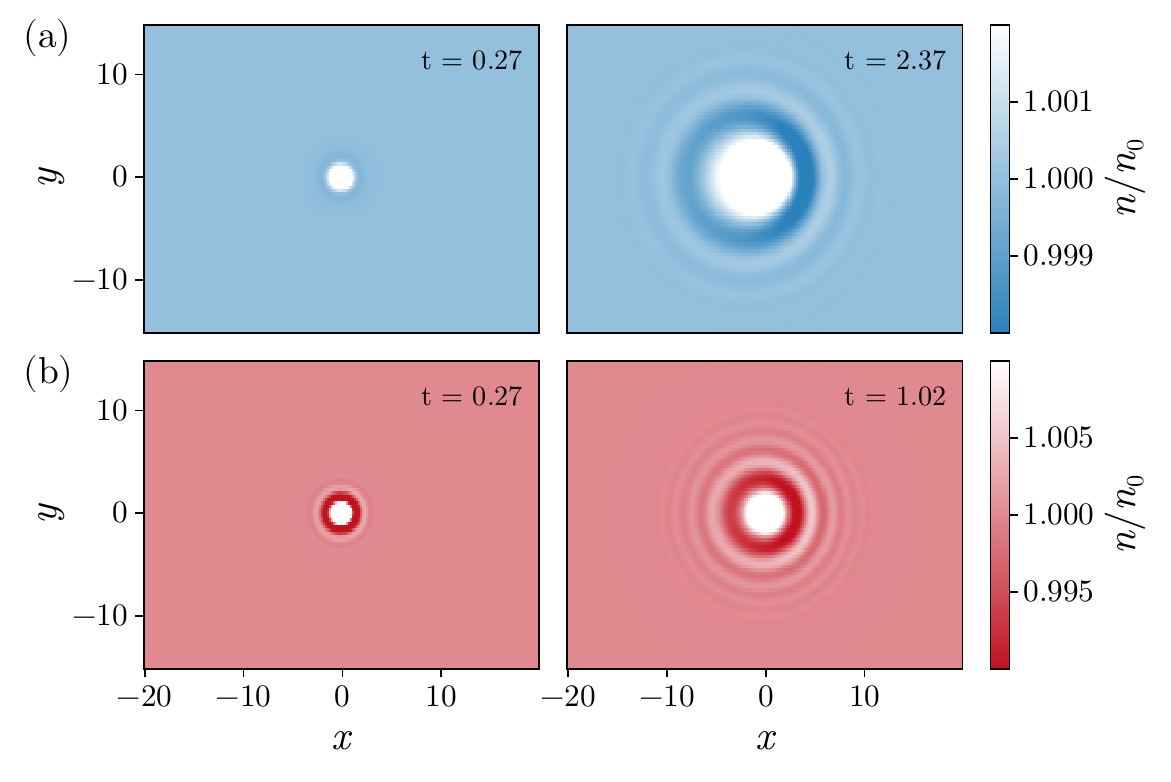}
\caption{\label{fig:dynamics} Snapshots of the planar cut through the condensate density at two different times for the ion-atom (upper) and gaussian (lower) potentials for interaction strengths $C_4=0.28$, $V_0=1.08$}
\end{figure}
\begin{figure}
\includegraphics[width = .98\columnwidth]{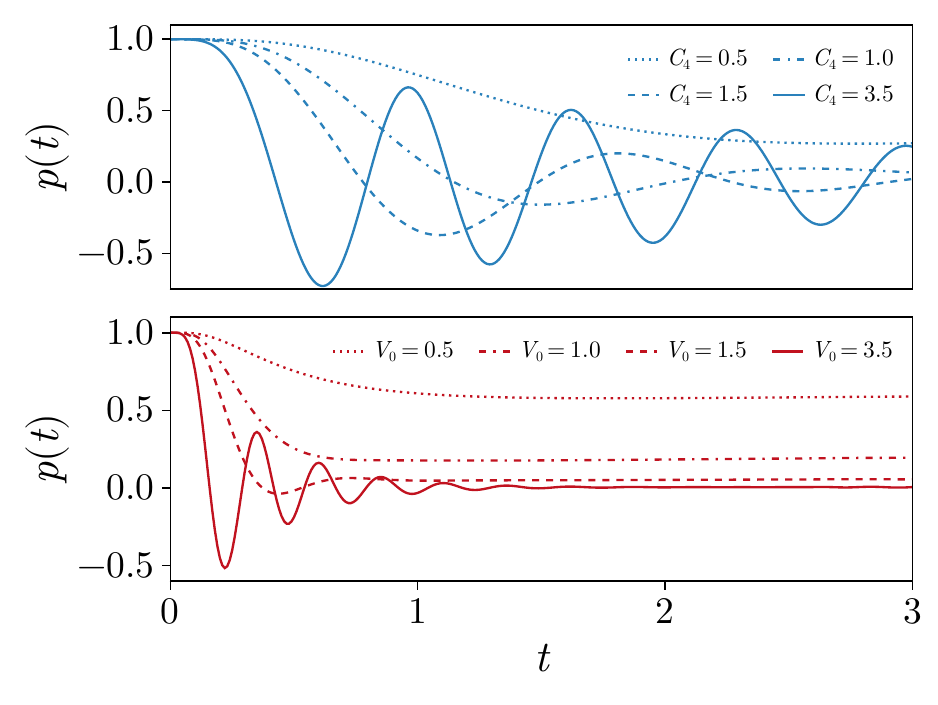}
\caption{\label{fig:osc} Momentum of the impurity as a function of time for the ion-atom (upper panel) and gaussian (lower panel) case for several potential strength values, showing the onset of decaying oscillations for sufficiently large interaction.}
\end{figure}
The polaron residue (quasiparticle weight) is an important property with respect to many-body nature of the interactions. It is defined as the overlap of the asymptotic wave functions of the polaron with the non-interacting one, $Z=|\langle \psi(t\to\infty)|\psi_0 \rangle|$, and can be calculated as~\cite{Yogurt2025}
\begin{equation}
	Z=e^{-\frac{1}{2}\sum_{\mathbf{k}} |\delta\psi_\mathbf{k}|^2}\, ~,
\end{equation}
where $\delta\psi_\mathbf{k}$ is obtained by decomposing the wave function in the Fourier space (for no interactions the condensate profile is flat so all its higher Fourier components are zero). The calculated residues are presented in Figure~\ref{fig:residue}. For both potentials they monotonically decay to zero with increasing interaction strength. For finite total momentum the residues in the stationary state have slightly lower values, but it does not lead to any qualitative changes. We can see that as the interaction strength grows, the gas density profile deviates more and more from the constant $n_0$. 

The wave-function overlap is a global property of the system, so in order to probe the short-range correlations more specifically, we study the Tan contact~\cite{Tan2008,Tan2008b,Guenther2021,Schmidt2022}. It is defined as a variation of the Gross-Pitaevskii energy functional with respect to the inverse of the impurity-boson scattering length $a_{\mathrm{IB}}$, 
\begin{equation}
	C = -\frac{8\pi m_{r}}{\hbar^2}\frac{\partial\mathcal{E}_{0}}{\partial(1/a_{\mathrm{IB}})} ~.
\end{equation}
In  Fig.~\ref{fig:tan-contact}, the Tan contact is depicted for three scenarios. First, within the mean field zero-range approximation it is described solely by the scattering length with a singularity at $1/a_{\mathrm{IB}}= 0$ (black dotted line). For finite-range interactions the divergence does not occur, and the maximum is expected to shift to the repulsive side of the resonance~\cite{Schmidt2022} (green dashed line). With the chosen parameters, power-law interaction produces a sharp maximum, located at positive scattering length rather close to the resonance position (red solid line). What might seem surprising is that the gaussian potential leads to a wider and lower peak at larger $a_{\mathrm{ia}}$. It may seem intriguing that in this particular case, the long-ranged potential yields results which are closer to the zero-range scenario. This is, however, in agreement with the results reported in~\cite{Schmidt2022}. Indeed, the contact is designed to characterize the short-range impurity-boson correlations which in our example are more pronounced for the gaussian case due to much higher boson density at the origin, as can be seen in Fig.~\ref{fig:ground}. Therefore, both the residue and contact are strongly connected with the peak density of the gas near the impurity.

Next, we calculate the effective mass of the polaron by fitting the energy of the stationary state solution at small momenta to parabolic dispersion $E=\frac{p^2}{2m^\star}$. The results are shown in Fig.~\ref{fig:meff} and again turn out to be qualitatively similar for the two potentials: short- and long-range interactions, especially at weak coupling strength where Born approximation is expected to work and polaron physics becomes universal. However, quantitative differences between the studied cases are still present. For the parameters used in data presented in Fig.~\ref{fig:ground}, the gaussian polaron effective mass is increased only by a few percent, while the charged, long-range interacting polaron mass is already almost twice the bare mass. This shows that for comparable dressing cloud size $N_{\rm cloud}$ the effective mass can already differ, contrary to the naive expectation, that $m^* - m \propto \Delta N$. We can therefore infer that not all atoms contribute to the mobility of the quasiparticle even though the polaronic cloud moves through the medium as a whole. The two quantities can nevertheless be expected to be correlated. Clearly, the equilibrium polaron properties in the studied regime are sensitive not only to the scattering length, but also the range of the interactions. In the next section, we explore the much more prominent differences in the dynamics.

\subsection{Relaxation dynamics}
\begin{figure}
	\includegraphics[width = .98\columnwidth]{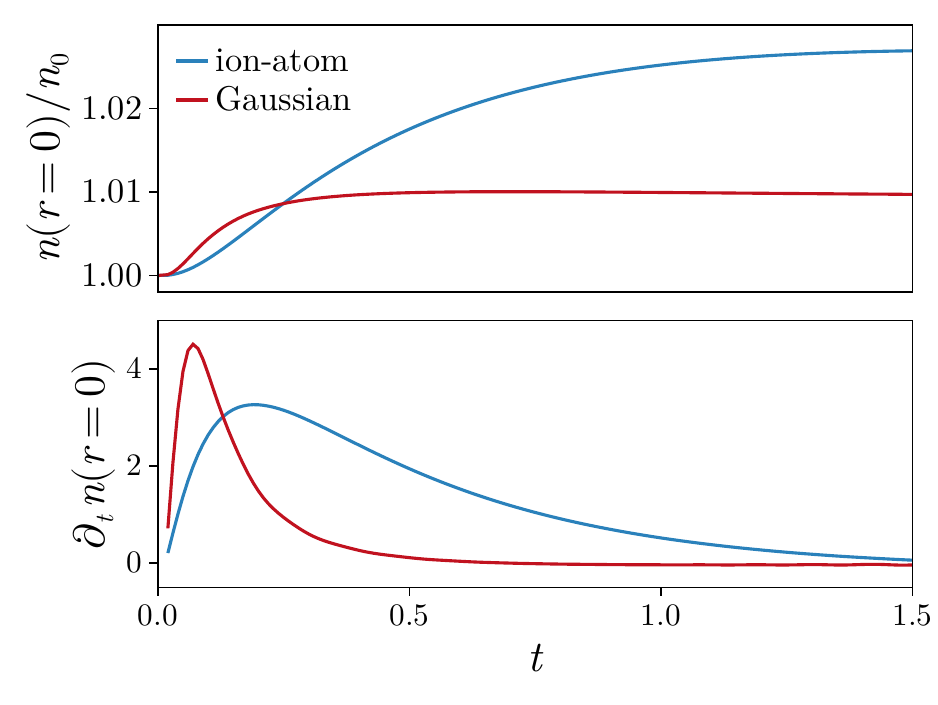}
	\caption{\label{fig:formation} Upper: buildup of the central density for an initially non-interacting system for ion-atom and gaussian potentials with $C_4 = 0.28$ and $V_0 = C_4/b^4 \approx 0.055$. Lower: time derivative of the central density for the same two cases.}
\end{figure}
\begin{figure}
 \includegraphics[width = .98\columnwidth]{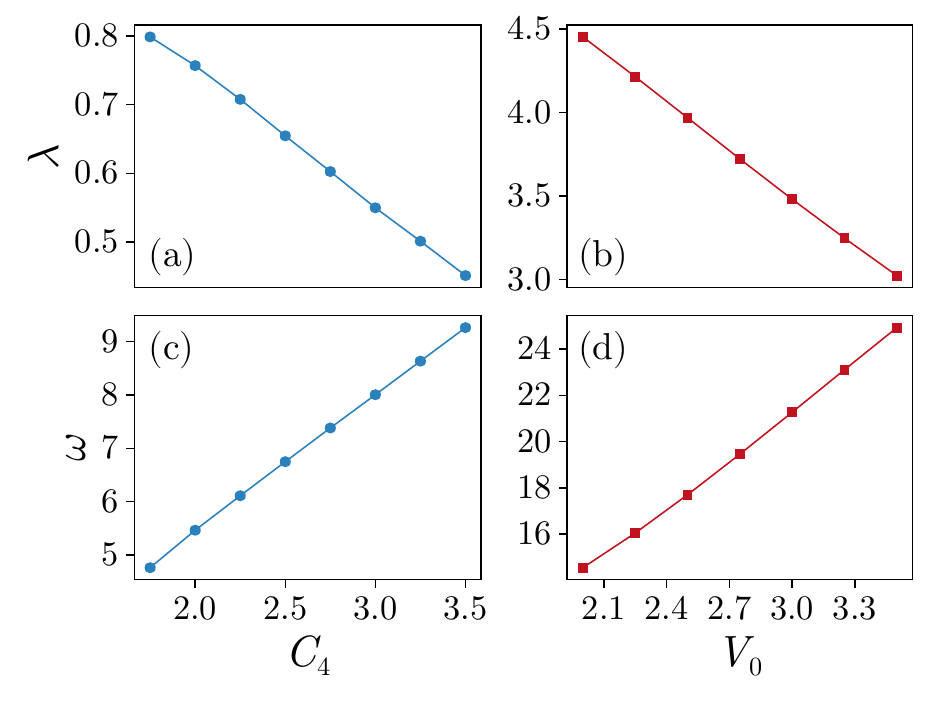}
	\caption{\label{fig:fits}Characterization of damped oscillations of the impurity velocity as a function of the interaction strength for ion-atom (left column) and gaussian(right) interactions. Upper row: damping rate $\lambda$, lower row: oscillation frequency $\omega = \sqrt{\omega_0^2 - \lambda^2}$.}
\end{figure}
We now consider a specific dynamical protocol in which the impurity is injected into the homogeneous gas with a finite initial velocity. This scenario can be realized by an electric field pulse giving the ion a momentum kick~\cite{Dieterle2021}, or by creating the impurity in the gas by means of a two-photon Raman process, where the recoil momentum can be controlled by tuning the angle between the beams~\cite{Grimm2025}. The evolution of the density profile of the condensate for both interactions is shown in Fig.~\ref{fig:dynamics} for $|{\bm p}_0| = 1$ in our units. We clearly observe that the dynamics is complex and the gas gets visibly excited despite the fact that the impurity velocity is always smaller than the speed of sound, similarly to what we already presented in Ref.~\cite{Wysocki2026} for a long-range interacting impurity. Note that in the vicinity of the impurity the speed of sound  is locally increased as it is proportional to the density, adding additional layer of complexity. Furthermore, the emitted phonons move in a finite range interaction potential, meaning that their propagation will be strongly affected for long-range interactions. The anisotropy of the excited waves is due to the fact, that the initial momentum of the impurity breaks the rotational symmetry of the system. After some evolution time, the impurity asymptotically reaches the stationary state.

At this point, one may already notice that the dynamics of the long-range interacting impurity is vastly different than for the short-range potential. This is best illustrated by tracking the impurity momentum in time, as shown in Fig.~\ref{fig:osc}. The stationary value of the impurity momentum depends on the interaction strength, and typically differs from zero but decreases with increasing coupling strength. Intuitively, stronger interactions lead to more excitations, and as the emitted phonons carry away the energy, the impurity has to decelerate more. It is also important to notice that for the two interactions the stationary state is attained on very different timescales. Namely, the ion-atom potential is characterized by much longer relaxation time with respect to the gaussian potential. In order to better understand how long it takes to reach the stationary state, we analyze the density buildup at the center after suddenly injecting the impurity into the gas. In Figure~\ref{fig:formation} we observe that the maximum time derivative of the density, marking the fastest growth, occurs later for the power-law potential and decays more slowly. Note that in this example we have adjusted the gaussian potential to lead to a similar peak gas density as obtained for the ion-atom case.

Finally, Fig.~\ref{fig:osc} shows oscillations of the impurity momentum, which appear at strong coupling. This phenomenon was observed in previous works~\cite{Majumdar2025,Wysocki2026} and was recently described in detail in the one-dimensional case~\cite{Majumdar2026}. The oscillations are ubiquitous and appear in all dimensions and for any interaction type. They can be understood from the hydrodynamic set of equations~\eqref{eq.hydro1}-\eqref{eq.hydro2} introduced earlier. Equation~(\ref{eq.dampedho}) provides an excellent fit to the non-linear impurity dynamics for both potentials. Furthermore, the parameters feature a linear dependence on the interaction strength in the probed regime, as shown in Fig.~\ref{fig:fits}. We have also verified that the oscillation period does not depend on initial impurity momentum. We leave a detailed analytical calculation of the parameters $\lambda$ and $\omega$ for future work.

\section{Conclusions}
\label{sec:concl}
We have analyzed the dynamics of an impurity immersed in a weakly interacting Bose gas, focusing on the role of the interaction range. Within the mean field theory applied in the co-moving frame, we observed slowing down of the impurity and possible damped oscillatory motion at large coupling strength. The oscillations have universal features regardless of the interaction type, and simply depend on the overall magnitude of the coupling. However, the interaction range has a very strong impact on the characteristic timescales, leading to much slower equilibration for power-law interactions. Furthermore, the effective mass in not increased proportionally to the number of bath atoms participating in the polaron cloud, and features a nonuniversal behavior. These predictions should be applicable to experiments with moving Bose polarons. 

\bigskip
\section{Acknowledgments}
This work was supported by the National Science Centre, Poland (NCN), Contract No. 2020/37/B/ST2/00486 (P.W., U.C.O., K.J.). U.C.O. acknowledges financial support from the National Council for Humanities, Sciences and Technologies of Mexico (CONAHCYT, now SECIHTI) through a doctoral scholarship (Support No. 4027346). We gratefully acknowledge Polish high-performance computing infrastructure PLGrid (HPC Centers: ACK Cyfronet AGH) for providing computer facilities and support within computational grant no. PLG/2025/018416. 

\bibliography{refs}

\end{document}